\documentclass[12pt]{amsart}
\usepackage{amsmath}
\newcommand{\cal}{\mathcal}
\newcommand{\lh}{{\cal L}({\cal H})}
\newcommand{\hi}{{\cal H}}
\newcommand{\ip}[2]{\left\langle\,#1\,|\,#2\,\right\rangle}
\newcommand{\ket}[1]{\mid#1\rangle}
\newcommand{\tr}[1]{{\rm tr}[#1]}
\newcommand{\kb}[2]{|#1\,\rangle\langle\,#2|}
\newcommand{\fii}{\varphi}
\newcommand{\R}{\mathbb R}
\newcommand{\br}{\mathcal B(\mathbb R)}

\begin{document}
\title[A dilemma in quantum mechanics]{A dilemma in representing observables\\
in quantum mechanics}
\author{D. A. Dubin}
\address{Daniel A. Dubin, Department of Pure Mathematics, The Open University, Milton Keynes, MK7 6AA, England}
\email{d.a.dubin@open.ac.uk}
\author{M. A. Hennings}
\address{Mark Hennings, Sidney Sussex College, Cambridge, CB2 3HU, England}
\email{mah30@cam.ac.uk}
\author{P. Lahti}
\address{Pekka Lahti, Department of Physics, University of Turku,
20014 Turku, Finland}
\email{pekka.lahti@utu.fi}
\author{J.-P. Pellonp\"a\"a}
\address{Juha-Pekka Pellonp\"a\"a, Department of Physics, University of Turku,
20014 Turku, Finland}
\email{juhpello@utu.fi}
\date{\today}
\begin{abstract}
There are self-adjoint operators which  determine both spectral and
semispectral measures. These measures have very different commutativity and
covariance  properties. This fact poses a serious question on the physical
meaning of such a self-adjoint operator and its associated operator measures.
\end{abstract}
\maketitle

\maketitle

\section{Introduction}
It is well-known that a given self-adjoint  operator may occur
as the first moment operator of various semispectral measures,
including its unique spectral measure. It is, perhaps,  less
widely known that there are self-adjoint operators which
uniquely  determine  not only their spectral measures but also
their semispectral measures. This situation seems to pose a
dilemma in the traditional text book wisdom of quantum mechanics
whereby physical quantities, also called observables, are
represented as self-adjoint operators.

In this note we wish to draw attention to this dilemma by means of
examples. To avoid an early  commitment to a particular approach
to quantum observables, in the main body of the paper we use the
standard mathematical terminology of self-adjoint operators,
spectral measures,  and semispectral measures. The discussion on
the physical meaning of the mathematical formalism is postponed
till the final section of the paper.

The use of semispectral measures (normalized positive
operator-valued measures) both in analysing actual experiments and
in  studying conceptual and mathematical foundations of quantum
mechanics has increased greatly during the last three decades, as
can be seen by the appearance of a number of  monographs on the
subject, \emph{q. v.}
\cite{Davies,Holevo,Ludwig,BLM,Peres,BGL,Alibook,Holevo1} amongst
others.

Though not exclusively, the need to use  semispectral instead of
spectral measures (projection-valued measures) is often
explained, explicitly or implicitly, as resulting from some
uncontrollable aspects or statistical decisions. The results of
this paper support the view that semispectral measures have a
fundamental r\^{o}le in quantum mechanics beyond this. For one
thing, a semispectral measure can be assigned to observables
whose `spectrum' is, say, a curved surface (of moderate
regularity) which is considerably more difficult to describe in
purely operator theoretic terms.

\section{Description of the problem}
Let $A$ be a self-adjoint operator, with a domain of definition $\mathcal
D(A)$, and let $E:\br\to\lh$ be a spectral measure, defined on the Borel
subsets of the real line $\R$ and taking values on the set $\lh$ of bounded
operators on $\hi$.

For any two vectors $\fii,\psi\in\hi$ we let $E_{\fii,\psi}$ denote the complex
measure $X\mapsto E_{\fii,\psi}(X) :=\ip{\fii}{E(X)\psi}$. According to the
spectral theorem for self-adjoint operators, any spectral measure $E$
determines a unique self-adjoint operator $A$, with the domain $\mathcal D(A)$,
such that for any $\fii\in\hi,\psi\in\mathcal D(A)$,
\begin{eqnarray}
\ip{\fii}{A\psi} &=& \int_{\R} x\, dE_{\fii,\psi}(x) \label{E1}\\ \mathcal D(A)
&=& \{ \psi\in\hi \,|\, \int_\R x\, dE_{\fii,\psi}(x) \ {\rm exists\ for\ all}\
\fii\in\hi \} \\ &=& \{ \psi\in\hi\,|\, \int_{\R} x^2 \, dE_{\psi,\psi}(x)<
\infty \},\label{da}
\end{eqnarray}
and, conversely, any self-adjoint operator $A$ determines a unique spectral
measure $E$ such that the above relations are valid. We let $E^A$ stand for the
spectral measure of $A$, and we note that $A$ is the first moment operator of
the operator measure $E^A$.

Due to the multiplicativity of the spectral measure, the $k$-th
moment operator $E^A[k]$ of $E^A$ is the $k$-th power of its
first moment operator $A$. That is, for any $k\in\mathbb N$, $$
E^A[k] :=\int_\R x^k\, dE^A(x) = (\int_\R x\, dE^A(x))^k =
(E^A[1])^k = A^k, $$ where the operator equalities are in the
weak sense, as in  (\ref{E1}), with a definition for $\mathcal
D(A^k)$ analogous to that given in (\ref{da}). It is also well
known that the spectrum of $A$, $\sigma(A)$, is equal to the
support, ${\rm supp}\, (E^A)$, of $E^A$: $$ \sigma(A)= {\rm
supp}\, (E^A). $$

In two recent papers \cite{LP00,DH00} it was independently shown
that there are self-adjoint operators $A$ which both uniquely
determine and are determined by certain semispectral measures
$F:\br\to\lh$ such that for all $\fii\in\hi,\psi\in\mathcal D(A)$,
\begin{eqnarray}
\ip{\fii}{A\psi} &=& \int_\R x\, dF_{\fii,\psi}(x),\\ \mathcal D(A) &=& \{
\psi\in\hi \,|\, \int_\R x\, dF_{\fii,\psi}(x) \ {\rm exists\ for\ all}\
\fii\in\hi \} \\ &\supseteq& \{\psi\in\hi\,|\, \int
x^2\,dF_{\psi,\psi}(x)<\infty \}.\label{inclusion}
\end{eqnarray}
In other words, $A$ is the first moment operator of the operator
measure $F$ and $F$ is uniquely determined by $A$. Since $F$ is
not the spectral measure the set inclusion (\ref{inclusion}) may,
in general, be a proper one.\footnote{We recall that it is the
positivity of the operator measure $F$, that is, $F(X)\geq O$ for
all $X\in\br$, which implies that the square integrability domain
of Eq. (\ref{inclusion}) is a subspace of $\mathcal D(A)$. If, in
addition, $F$ were projection valued, that is, $F(X)=F(X)^2$ for
all $X\in\br$, then the set inclusion (\ref{inclusion}) would be
an equality \cite[Lemma A.2]{LMY98}, c. f. Eq. (\ref{da}) }

The self-adjoint operators $A$ in question are of a special type,
specifying and being specified by semispectral measures $F$ with
particular additional properties. In spite of this mutual
specification, however, the $k$-th moment operator will not be the
$k$-th power of the first moment operator in general. In
particular, it will be the case that
$$ F[2] = \int x^2\,dF(x) \ \geq \  ( \int x\,dF(x))^2 = F[1]^2
=A^2. $$ It must be stressed that it is quite exceptional for a
semispectral measure $F$ to be determined by its first moment
operator $F[1]=A$. For a general semispectral measure $F$, even
the knowledge of the moment operators $F[k]$ for all $k\in\mathbb
N$ will not suffice to determine $F$.

However, if the support of $F$ is compact then its moment
operators $F[k], k\geq 0$, are bounded self-adjoint operators and
the operator sequence $F[k], k\geq 0$, determines the operator
measure $F$.\footnote{This is a well-known consequence of the
Weierstrass approximation theorem and the uniqueness part of the
Riesz representation theorem.} Clearly, the spectrum of $A$ is
then a subset of the support of $F$, $$ \sigma(A)\subseteq {\rm
supp}\,(F), $$ with the possibility that the inclusion is a
proper one.

Given this background, we now state the conundrum for quantum
theory that we referred to above.

According to the usual text book formulation of quantum mechanics
physical quantities are represented by self-adjoint operators,
and, usually, even the converse is assumed (if no superselection
rules are involved): each self-adjoint operator corresponds to a
physical quantity. The mathematics just described raises the
following question: if a given self-adjoint operator $A$ gives
rise to a unique spectral measure $E^A$ and (by the formula
prescribed in \cite{LP00,DH00}) a unique semispectral measure $F$,
and these two measures are not the same, $E^A\ne F$, what is the
relationship of the operator $A$ and the two measures $E^A$, $F$,
to the observable? If the observable is represented by the
self-adjoint operator $A$, what is to be understood by the
differing measure representations $E^A$, $F$, that it has? If, on
the other hand, observables are completely represented by
measures, then what is to be made of the fact that the
self-adjoint operator $A$ is now associated with two distinct
observables, $E^A$and $F$? We shall investigate these questions by
considering three sets of examples.

The first two of them give  examples of the situation described
above, whereas the third illustrates the commonly accepted
viewpoint  that  some semispectral measures associated with a
self-adjoint operator are to be interpreted as smeared, noisy,
unsharp, or inaccurate versions of the observable represented by
the spectral measure of the self-adjoint operator.

\section{Examples}
The examples that illustrate the problem   arise from the theory of generalized
imprimitivity systems, called also systems of covariance. The primary examples
were discussed in \cite{LP00, DH00}. Here we follow \cite{CDVLP02} and
\cite{CDV,CDVpri} to provide somewhat wider classes of relevant examples.

\subsection{$\mathbb N$-covariant semispectral measures}\label{Ncov}
Let $\hi$ be a complex separable Hilbert space,  $\{\ket n\}_{n\in\mathbb N}$ a
fixed  orthonormal basis for $\hi$, and let $N$ denote the self-adjoint
operator for which $N\ket n = n\ket n$ for all $n\in\mathbb N$.

Consider a semispectral  measure $F$ defined on the Borel subsets
of the interval $[0,2\pi)$ and taking values in $\lh$. We say
that $F$ is $\mathbb N$-covariant if it forms together with the
unitary representation $x\mapsto e^{ixN}, x\in\mathbb R$, of the
additive group of $\mathbb R$, a generalized imprimitivity system,
that is,
\begin{equation}\label{covariance}
e^{ixN}F(X)e^{-ixN} = F(X+x)
 \end{equation}
for all $X\in \mathcal{B}([0,2\pi)), x\in\mathbb R$, where the
addition $X+x$ is modulo $2\pi$. (The labelling of the covariance
by $\mathbb N$ refers to the spectrum of $N$.)

The structure of the $\mathbb N$-covariant semispectral measures
$F: \mathcal{B}([0,2\pi))\to\lh$ is well known,  see e.g.
\cite{Holevo, LP99}. Perhaps, the simplest way to characterize
them is the following \cite{CDVLP02}: $F$ satisfies the
covariance condition (\ref{covariance}) if and only if there is a
(not necessarily orthogonal) sequence of unit vectors
$(h_n)_{n\in\mathbb N}$ of $\hi$ such that for any
$X\in\mathcal{B}([0,2\pi))$,
\begin{equation}\label{fx}
F(X) = \sum_{n,m\in\mathbb N} \ip{h_n}{h_m}\, \tfrac{1}{2\pi}\int_Xe^{i(n-m)x}d\,x\,\kb nm,  
\end{equation}
where the series converges weakly.

It is to be noted that two sequences of unit vectors $(h_n)_{n\in\mathbb N}$
and $(h'_n)_{n\in\mathbb N}$ determine the same semispectral measure $F$
exactly when $ \ip{h_n}{h_m}= \ip{h'_n}{h'_m}$ for all $n,m\in\mathbb N$.

By a direct computation one may easily confirm that the only
commutative solution of (\ref{covariance}) is the scalar measure:
the commutativity  of $F$, that is, $F(X)F(Y)=F(Y)F(X)$ for all
$X,Y\in\mathcal{B}([0,2\pi))$, holds if and only if the generating
vectors $h_n$ are pairwisely orthogonal; in that case $F(X)
=\tfrac 1{2\pi}\int_Xd\,x\, I$ for all
$X\in\mathcal{B}([0,2\pi))$.\footnote{For a full analysis of the
degree of commutativity of the $\mathbb N$-covariant semispectral
measures $F$, see \cite{BLPY01}.} In particular, this means that
among the $\mathbb N$-covariant semispectral measures (\ref{fx})
there is no spectral  measure.

The  $\mathbb N$-covariant semispectral measures $F:{\cal
B}([0,2\pi))\to\lh$ are supported by the  interval $[0,2\pi]$.
Therefore, their moment operators,
\begin{eqnarray}
F[k] &=& \int_0^{2\pi} x^k \, dF(x),\  k\in\mathbb N, \label{fk}\\
&=&\sum_{n,m\in\mathbb N}\ip{h_n}{h_m}\, \tfrac
1{2\pi}\int_0^{2\pi}x^ke^{i(n-m)x}\,dx\,\kb nm\, , \nonumber
\end{eqnarray}
are bounded self-adjoint operators. Since no such $F$  is
projection valued, as noted above, the second moment operator
$F[2]$ is never equal to the square of the first moment operator
$F[1]$,
\begin{equation}
F[2] \geq F[1]^2,\   \  \ F[2] \ne F[1]^2,
\end{equation}
see, for instance, \cite[Appendix, Eq. 9 on p. 446]{SZNagy}.

The covariance condition (\ref{covariance}) completely determines
the structure of the semi\-spectral measures (\ref{fx}) and thus
also their moment operators (\ref{fk}). But from (\ref{fk}) we
see that
\begin{equation}
\ip{h_n}{h_m} = \ip{n}{F[1]|m}i(n-m),\ n\ne m,
\end{equation}
which implies that any  $\mathbb N$-covariant semispectral measure $F$ is
uniquely determined by its first moment operator $F[1]$,
\begin{equation}\label{firstmoment}
F[1] = \pi I+\sum_{n\ne m\in\mathbb N}\frac{\ip{h_n}{h_m}}{i(n-m)}\,\kb nm,
\end{equation}
see \cite{LP00, DH00}.

As a bounded self-adjoint operator,  $F[1]$
 has a unique spectral measure $E^{F[1]}$ such that
\begin{equation}
F[1] = \int_{\mathbb R} x\,dE^{F[1]}(x),
\end{equation}
where the integral ranges effectively over the spectrum of $F[1]$. The operator
$F[1]$ determines thus both a unique $\mathbb N$-covariant semispectral measure
$F$ and a unique spectral measure $E^{F[1]}$. A distinctive feature is that the
spectral measure $E^{F[1]}$ cannot be $\mathbb N$-covariant. Also, apart from
the trivial case, $F$ is noncommutative whereas $E^{F[1]}$ is multiplicative
and thus commutative. We note further that ${\rm supp}\, (E^{F[1]}) =
\sigma(F[1]) \subseteq {\rm supp}\,(F)$.

Perhaps the most natural and important example of the $\mathbb
N$-covariant semispectral measures $F$ is the one associated with
a constant sequence $h_n=h$, $n\in\mathbb N$. (Any unit vector
will do; see above.) This semispectral measure has been advocated
by some authors as the canonical phase observable $F_{{\rm can}}$,
see e.g. \cite{Holevo,BGL,LP00}. However, since its first moment
is not canonically conjugate to the number operator, employing the
word ``canonical" in this context is not the familiar textbook
usage; canonicity here is with respect to the above class of
semispectral measures.

Similarly, its first moment operator\footnote{This operator is
unitarily equivalent to the Toeplitz operator of multiplication
by the independent variable on the Hardy Hilbert subspace of
square integrable functions on the circle.}
\begin{equation}\label{GW}
F_{{\rm can}}[1]= \pi I + \sum_{n\ne m=0}^\infty\frac 1{m-n}\,\kb nm,
\end{equation}
is frequently proposed as  the phase operator, see e.g.
\cite{GW,Calindo,Mlak}. In this case $\sigma(F_{{\rm
can}}[1])={\rm supp}\,(F_{{\rm can}})$. The spectral measure of
$F_{{\rm can}}[1]$ has a rather complicated structure, see
\cite{GW} for an analysis; nevertheless it is  not $\mathbb
N$-covariant. This and other candidate phase observables are
discussed at length in \cite{DHS}.

\subsection{$\mathbb Z$-covariant semispectral measures}\label{Zcov}
Taking an orthonormal basis labelled by the set of all, and not
simply non-negative, integers leads to a class of examples
similar to those obtained in Section \ref{Ncov}. Some new
 and interesting features do arise with this choice, however. Therefore,
let$\{\ket k\}_{k\in\mathbb Z}$ be an orthonormal basis of $\hi$,
and let $Z$ denote the self-adjoint operator with $Z\ket k =
k\ket k$ for all $k\in\mathbb Z$.

Extending the previous terminology, we say that a semispectral
measure $F: \mathcal{B}([0,2\pi))\to\lh$ is $\mathbb Z$-covariant
if it satisfies the covariance condition
\begin{equation}\label{Zcovariance}
e^{ixZ}F(X)e^{-ixZ} = F(X+x)
\end{equation}
for all $X\in \mathcal{B}([0,2\pi)), x\in\mathbb R$, where the addition $X+x$
is modulo $2\pi$.

As  in Section \ref{Ncov}, a semispectral measure $F$ is $\mathbb Z$-covariant
if and only if there is a sequence of unit vectors $(h_k)_{k\in\mathbb Z}$ of
$\hi$ such that for any $X\in\mathcal{B}([0,2\pi))$
\begin{equation}\label{zfx}
F(X) = \sum_{k,l\in\mathbb Z} \ip{h_k}{h_l}\, \tfrac{1}{2\pi}\int_Xe^{i(k-l)x}d\,x\,\kb kl. 
\end{equation}
The principal difference here is that among the solutions
(\ref{zfx}) of the covariance condition (\ref{Zcovariance}) there
are both commutative and noncommutative semispectral measures
and, in addition, a spectral measure (unique up to unitary
equivalence) obtained with an arbitrary choice of unit vector
$h_k=h$ for all $k\in\mathbb Z$. For each solution $F$, the first
moment operator $F[1]$ uniquely determines the semispectral
measure $F$.\footnote{The structure of the moment operators are
like in the $\mathbb N-$covariant case, with the sole  exception
that now the summations are over $\mathbb Z$, \emph{c.f.}
\cite{CDVLP02}.}

Clearly, the spectral measure $E^{F[1]}$ of $F[1]$ is $\mathbb Z$-covariant
exactly when it coincides with $F$, which is now the case for the constant
sequence $h_k=h$, $k\in\mathbb Z$. In this case the pair $(F_{{\rm can}}[1],Z)$
constitutes  a Schr\"odinger pair, that is, the usual position-momentum
operators of a particle in a box of length $2\pi$.

\subsection{$\mathbb R$-covariant semispectral measures}\label{Rcov}
To emphasize the very special nature of the previous two sets of
examples, let us consider next the multiplicative  operator $Q$ in
$L^2(\mathbb R)$, $(Q\fii)(x)=x\fii(x)$, with the domain
$\mathcal D(Q) =\{\fii\in L^2(\mathbb R)\,|\, \int_{\mathbb R}
x^2|\fii(x)|^2\,dx < \infty\}$. 
Consider the unitary representation $x\mapsto U_x$ of the real
line $\mathbb R$ given by $(U_x\fii)(y) =\fii(y-x)$. As well
known, the spectral measure of $Q$, $E^Q$, is  (up to unitary
equivalence) the unique projection-valued solution of the $\mathbb
R$-covariance condition \cite{Mackey}
\begin{equation}\label{rcov}
U_xF(X)U_x^* = F(X+x), \ \ \ X\in\br,\ x\in\mathbb R \,.
\end{equation}
However, this covariance condition (\ref{rcov}) can be solved for
arbitrary semispectral measures \cite{Holevo}, and one obtains
thereby both commutative and noncommutative semispectral measures
in addition to the spectral measure \cite{CDVpri}. In particular,
any convolution of $E^Q$ with a probability density $f$ yields an
$\mathbb R$-covariant semispectral measure of the form
$E^{Q,f}(X) =(\chi_X*f)(Q)$, $X\in\br$, $f=|\eta|^2,\eta\in
L^2(\mathbb R)$, $\parallel\eta\parallel =1$. If the expectation
value of the density function $f$ is zero, then the first moment
operator of $E^{Q,f}$ equals the first moment operator of $E^Q$,
namely $Q$, see, e.g. \cite{Davies}. Therefore, $Q$ cannot
determine $E^{Q,f}$, so the conundrum posed above does not occur
in this case.

\section{Discussion}
In the approach to quantum mechanics which starts with the
operational idea of a preparation and registration procedure one
is lead in a natural way to the set of states and the set of
observables being in duality. The states may be defined as
equivalence classes of preparations, the observables as
totalities of measurement outcome statistics.

States may be represented as positive normalized trace class
operators $\rho$ (known as density operators) on the
configuration Hilbert space $\hi$ for the system under
consideration. An observable may then be defined as a normalized
semispectral measure $F$ defined on the relevant
($\sigma$-algebra of subsets of) space of values (measurement
outcomes), typically the Borel $\sigma$-algebra $\br$ of the real
line $\mathbb R$, see  e.g. \cite{Davies, Holevo, Ludwig, BGL}.
The probability measure $X\mapsto F_\rho(X):=\tr{\rho F(X)}$,
defined by a state $\rho$ and an observable $F$, is then taken to
describe the measurement outcome statistics obtained when the
same $F$-measurement is repeated under the same conditions,
described by $\rho$, a large number of times.

In this approach, spectral measures appear as special idealized
cases, called decision observables in \cite{Ludwig}, ordinary
observables in \cite{BLM},  and sharp observables in \cite{BGL}.
The first moment operator  $F[1]$ of an observable $F$ accounts
for the expectation value $\tr{\rho F[1]}$ of the probability
distribution $F_\rho$. The knowledge of the expectation values
$\tr{\rho F[1]}$, for all states $\rho$, determines the operator
$F[1]$, and thus its spectral measure $E^{F[1]}$ as well. In
general, it does not determine the semispectral measure $F$. But
the examples of Sections \ref{Ncov}-\ref{Zcov} show that there are
cases where this does happen.

To help to analyse the question of the physical meaning of the the
self-adjoint operator $F[1]$ and its spectral  measure $E^{F[1]}$
for $\mathbb N$-/$\mathbb Z$-covariant semispectral measures $F$,
we consider the resulting probability measures and their
variances.

From the probabilistic point of view, the spectral and
semispectral measures associated with $F[1]$ are quite different.
For while the probability distributions $F_{\fii,\fii}$ and
$E^{F[1]}_{\fii,\fii}$  in any vector state
$\fii\in\hi,\parallel\fii\parallel=1$, have the same expectations
\begin{equation}
\int_{\sigma(F[1])}x\,dE^{F[1]}_{\fii,\fii}(x) = \ip{\fii}{F[1]\fii} =
\int_0^{2\pi}x\,dF_{\fii,\fii}(x),
\end{equation}
their other moments are different. In particular, their variances
are different:
\begin{eqnarray}
{\rm Var}\,(F_{\fii,\fii}) &=& \int x^2\,dF_{\fii,\fii}(x)-(\int
x\,dF_{\fii,\fii}(x)^2 \\ &=&
\ip{\fii}{F[2]\fii}-\ip{\fii}{F[1]\fii}^2\nonumber \\ &=& \ip{\fii}{F[2]\fii}-
\ip{\fii}{F[1]^2\fii}+ \ip{\fii}{F[1]^2\fii}- \ip{\fii}{F[1]\fii}^2 \nonumber\\
&=&  \ip{\fii}{(F[2]-F[1]^2)\fii}+ {\rm Var}\,(E^{F[1]}_{\fii,\fii}) \nonumber
\\ &\geq& {\rm Var}\,(E^{F[1]}_{\fii,\fii}).\nonumber
\end{eqnarray}
This relation is sometimes taken to suggest that $F$ could be
regarded as  a smeared or noisy version of $E^{F[1]}$, since, in
a vector state $\fii$, the variance of $F$ is greater by a noise
term $\ip{\fii}{(F[2]-F[1]^2)\fii}$ than the variance of
$E^{F[1]}$.

A smearing of, or noisy version of, $E^{F[1]}$ is typically
obtained by convolving it with a density function $f$, that is, a
nonnegative Borel function $f:\mathbb R\to \mathbb R$, possibly
supported by $[0,2\pi]$, such that $\int f(x)\,dx=1$, see e.g.
\cite{AliDoeb,BGL}. The semispectral measure $E^{F[1],f}:X\mapsto
(\chi_X*f)(F[1])$ obtained in this way is commutative.\footnote{In
fact, any commutative semispectral measures can be represented as
a probability average of a unique spectral measure
\cite{Holevo72}, see also \cite[Sect. 2.1.3.]{Holevo1} and
\cite[Theorem 3.1.3]{Alibook}.}

All the (nontrivial) $\mathbb N$-covariant semispectral measures
of Section \ref{Ncov} are noncommutative. In contrast,
convolutions of spectral measures with probability densities are
commutative semispectral measures, and none of them is $\mathbb
N$-covariant. Note also that if the average of a smearing
function $f$ is zero, $\int xf(x)\,dx=0$, then the first moment
operator of $E^{F[1],f}$ is again $F[1]$, but, clearly, it cannot
determine $E^{F[1],f}$.

We conclude that the self-adjoint operators $F[1]$ of Section \ref{Ncov}
constitute  examples of self-adjoint operators which represent  different
observables $F$ and $E^{F[1]}$. Their measurement outcome statistics, described
by the probability measures $F_{\fii,\fii}$ and $E^{F[1]}_{\fii,\fii}$,   are
different, though they are indistinguishable by the statistical average. Their
difference becomes evident, for instance, in their standard deviations. From
the statistical point of view one may say that $E^{F[1]}$ is the observable
associated with $F[1]$ which has the least variance \cite{SZNagy}, whereas $F$
is the observable associated with $F[1]$ which is $\mathbb N$-covariant.

Among the solutions of $\mathbb Z$-covariant semispectral measures there are
also commutative measures, including the canonical spectral measure $F_{\rm
can}$. However, convolutions of $F_{\rm can}$ which have $F_{\rm can}[1]$ as
the first moment cannot be $\mathbb Z$-covariant. Therefore, also in this case
the (noncovariant) spectral measure and
the (covariant) semispectral measures 
of the self-adjoint moment operators $F[1]$ seem to represent
different, though in the sense of the statistical mean,
indistinguishable observables. It is also worth  noting that
among the commutative $\mathbb Z$-covariant semispectral measures
$F$ there is no smearing of $F_{{\rm can}}$ which would have the
same first moment as $F_{{\rm can}}$.

Finally, Section \ref{Rcov} gives examples of $\mathbb
R$-covariant semispectral measusures which can be interpreted as
smeared or unsharp versions of the sharp $\mathbb R$-covariant
spectral measure $E^Q$. In that case, however, the first moment
operator $Q$ of an $\mathbb R$-covariant
semispectral measure 
$E^{Q,f}$ does not suffice to determine the whole semispectral measure. We
recall from \cite[Theorem 3.3.2]{Davies} that if the density function $f$ has
finite mean and variance, then ${\rm Var}\,(E^{Q,f}_{\fii,\fii}) = {\rm
Var}\,(E^{Q}_{\fii,\fii}) +{\rm Var}(f)$ for any sufficiently smooth vector
states $\fii\in L^2(\mathbb R)$, that is, the noise term
$\ip{\fii}{E^{Q,f}[2]-E^{Q,f}[1]^2|\fii}$ is then simply ${\rm Var}(f)$.


\end{document}